# In-depth Analysis of Durations of Discretionary Lane Changes on Freeway Under Varying Traffic Conditions

Gen Li, Zhen Yang, and Yiyong Pan

*Abstract*—This paper aims to investigate the characteristics of durations of discretionary lane changes (LCs) on freeways based on an enriched dataset. A comprehensive analysis of LC durations was conducted based on vehicle types, LC directions and navigation speeds. It was found that the heavy vehicle takes longer time to complete LC maneuver. The LC direction significantly influences the durations of passenger cars but has no significant influence on heavy vehicles. The navigation speed was found to have important influence on LC durations. However, it has different impacts according to vehicle types and LC directions. Further analysis of LC durations at different stages showed that drivers of passenger cars might use different strategies to perform LCs when they change lanes to different directions. However, drivers of heavy vehicles in both directions used less time to occupy the target lanes. Results of this study can be beneficial to understand the mechanism of LC process and the influence of LC on traffic flow.

## I. INTRODUCTION

For quite a long time, LC studies were focused on decision modelling. A LC was considered as an instant event in many MTSs[1-3]. However, the LC process generally lasts for several seconds and the subject vehicle will interact with several surrounding vehicles during the process, making the LC execution one of the most dangerous driving behaviors. Some studies regarded the LC as a sequential decision process consisting of two steps: (i) decision making (ii) LC execution[4-8]. It was pointed by Jin [9] that the subject vehicle occupied two lanes during the LC process, increasing the traffic density and causing the capacity reduction. As a consequence, the LC duration becomes the most important factor that influences the range of capacity reduction. Thus, it is beneficial to investigate the LC duration to understand its influence on traffic flow, furthermore, improve the accuracy of MTSs. Unfortunately, only a few studies focused on the LC duration and even less studies emphasized the modelling of LC duration[8, 10, 11]. An important reason was the shortage of vehicle trajectory data. Though the commonly used NGSIM dataset has provided many insightful observations about LCs, the data were collected at weaving sections under congestion, which limits the generalization of the studies.

To understand the variation of LC duration under different traffic conditions, an in depth analysis of lane change durations based on an enriched naturalistic vehicle trajectory dataset collected on German highways by unmanned aerial vehicles (UAV) is conducted in this study. The dataset provides more than 110 500 vehicle trajectories under both free flow and congestion. Using more discretionary LC trajectories of 2905 passenger cars and 433 heavy vehicles under various traffic conditions, this study comprehensively investigates the discretionary LC duration of passenger cars and heavy vehicles. More specifically, this paper aims to achieve the following targets,

1) Explore the difference of LC durations between passenger cars and heavy vehicles,
2) Investigate the influences of traffic conditions on LC duration,
3) Built LC duration models for different vehicle types and LC directions.

## II. LITERATURE REVIEW

Since Gipps built a comprehensive LC framework based on gap acceptance theory in 1986[12], LCs has drawn more and more attention. Gipps' framework was widely used in several MTSs such as VISSIM, PARAMICS and CORSIM [13-15]. In these MTSs, different definitions of critical gaps were used to capture the LC decision behavior from different perspectives. Then, some studies reported the inconsistency of gap acceptance theory that vehicles might take LC actions when only the lead or lag gap, or even none of them, is larger than the critical gap [3, 7, 16-20]. Thus, discrete choice models such as logit and probit models were proposed to model LC decisions [21-25]. To better utilize the data, data mining techniques, such as CART, Bayesian Network and fuzzy logic models, were

This work was supported by Science and Technology Innovation Fund for Youth Scientists of Nanjing Forestry University under Grant CX2019021, Scientific Research Start-up Funds of Nanjing Forestry University under Grant GXL2020012 and Basic Research Program of Science and Technology Commission Foundation of Jiangsu Province under Grant BK20170932.
Li Gen is with College of Automobile and Traffic Engineering, Nanjing Forestry University, Nanjing, Jiangsu, P. R. China, 210037 (e-mail: ligen@njfu.edu.cn).
Zhen Yang is with College of Automobile and Traffic Engineering, Nanjing Forestry University, Nanjing, Jiangsu, P. R. China, 210037 (e-mail: zyang_2016@163.com).
Yiyong Pan is with College of Automobile and Traffic Engineering, Nanjing Forestry University, Nanjing, Jiangsu, P. R. China, 210037 (e-mail: uoupanyg@njfu.edu.cn).

applied to LC decision models[26-28].

However, most literatures focused on the LC decision but neglected the LC execution behavior, which might have a significant impact on traffic flow characteristics [8, 29-31]. Among the small literature on LC execution behaviors, some focused on the acceleration/deceleration behavior during the lane-changing process. Typical car-following models were modified to describe the LC execution behavior [30, 32-34]. To better understand the acceleration and deceleration behavior during LC process, a gradient boosting decision tree was applied to NGSIM datasets [8]. LC duration was also an important aspect of LC execution studies. LC duration is believed to have a significant impact on traffic flow characteristics[9, 35]. Therefore, LC duration also draws some attention from researchers. Different ranges of LC duration were reported in previous studies [10, 30, 36-38]. In general, the LC duration ranges from 1seconds to more than 10 seconds.

TABLE I
THE NUMBER OF DIFFERENT DLC SAMPLES

| Vehicle Type | LC Direction | Original Lane # | Target Lane # | Numbers | Total |
|---|---|---|---|---|---|
| Passenger Car | LCLL | 1 | 2 | 312 | 1355 |
| | | 2 | 3 | 1043 | |
| | LCRL | 2 | 1 | 452 | 1550 |
| | | 3 | 2 | 1098 | |
| Heavy Vehicle | LCLL | 1 | 2 | 198 | 213 |
| | | 2 | 3 | 15 | |
| | LCRL | 2 | 1 | 205 | 220 |
| | | 3 | 2 | 15 | |

However, limited studies focused on modelling the LC duration. Toledo and Zohar [10] first built LC duration models based on log-normal models by using the NGSIM dataset and found that LC durations were longer when the lane change vehicles faced more challenges. Cao, et al. [37] used a linear regression model based on 192 observations collected in Australia. However, the NGSIM dataset was collected under congestion and the dataset in Cao, et al. [37] was collected on an arterial road, which affected the generality of the findings.

In general, a number of investigations about the LC duration have been conducted in the literature. The results indicated that LC durations vary from one to another according to vehicle types and other factors. However, few researchers has modeled the durations of LCs. And the datasets used in those studies were limited in sample size. To fill the gap, an enrich dataset including 2905 passenger cars and 433 heavy vehicles under various traffic conditions is used in this study.

III. DATA PREPARATION

A. The highD Dataset

To analyze the LC behavior of heavy vehicles, the highway Drone (highD) dataset is chosen in this study. The data were recorded across German highways using drones. A drone with a high-resolution camera kept hovering at a fixed position above the highway to collect naturalistic driving data. As shown in Fig.1, the camera covered a section of highway of about 420 meters in length. The highD dataset contains 60 recordings with an average length of 17min at six locations of German highways with 2 or 3 lanes and different speed limits around Cologne captured during 2017 and 2018. Comparing with the commonly used NGSIM dataset, the highD dataset can provide much more data, especially for the heavy trucks. 20000 trucks were observed in the highD dataset, while the NGSIM dataset only have 278 trucks. Thus, this dataset is more suitable for analyzing the driving behavior of heavy vehicles. The highD dataset has also been used for car-following analysis by Kurtc [39] and achieved promising results. The speed analysis by Kurtc [39] showed that there was mostly free traffic in highD dataset, but it still contained a considerable datasets showing impeded traffic or even jams with stop-and-go waves because of its large sample size. For detailed information about the highD dataset, one can refer to Krajewski, et al. [40]. The highD dataset can be downloaded from *https://www.highd-dataset.com/*.

TABLE II
STATISTICS OF DURATION FOR DIFFERENT DIRECTIONS

| Vehicle Type | LC Direction | Sample Size | Mean (s) | Median (s) | Std. Dev (s) | Min (s) | Max (s) |
|---|---|---|---|---|---|---|---|
| Passenger Cars | Left | 1355 | 7.568 | 7.400 | 1.600 | 3.560 | 15.120 |
| | Right | 1550 | 7.785 | 7.560 | 1.629 | 4.320 | 21.600 |
| Heavy Vehicles | Left | 213 | 8.339 | 8.040 | 1.923 | 4.160 | 15.120 |
| | Right | 220 | 8.452 | 8.120 | 1.761 | 5.120 | 14.080 |

TABLE III
RESULTS OF MANN-WHITNEY-U TESTS FOR DIFFERENT VEHICLE TYPES AND LC DIRECTIONS

| Hypothesis | P-value | Results |
|---|---|---|
| Left LC vs Right LC for Passenger Cars | 0.001 | Reject $H_0$ |
| Left LC vs Right LC for Heavy Vehicles | 0.363 | Accept $H_0$ |
| Left LC for Passenger Cars vs Heavy Vehicles | 0.000 | Reject $H_0$ |
| Right LC for Passenger Cars vs Heavy Vehicles | 0.000 | Reject $H_0$ |

Among the three locations of highD dataset, the data collected at Location 1 are used in this study because of three reasons:

After these three steps, trajectories of 2905 passenger cars and 433 heavy vehicles were extracted from the dataset. Among them, 1350 passenger cars and 213 heavy vehicles change lanes to the left, while 1550 passenger cars and 220 heavy vehicles to the right. The distribution of different DLCs for passenger cars and heavy vehicles is shown in Table I. In Table I, the lane # 1, 2, 3 denote the rightmost, middle and leftmost lane, respectively. It can be found that most heavy vehicles made lane change in the middle and rightmost lane, while most passenger cars make lane change in the middle and leftmost lane.

TABLE IV
THE NUMBERS OF SAMPLES WITHIN DIFFERENT SPEED RANGES AND RESULTS OF MANN-WHITNEY-U TESTS FOR HEAVY VEHICLES

| Speed Range (m/s) | [0,20) | [20,25) | [25,30) | [30,35) | Total |
|---|---|---|---|---|---|
| Left | 28 | 101 | 75 | 9 | 213 |
| Right | 0 | 85 | 116 | 19 | 220 |
| Total | 28 | 186 | 191 | 28 | 433 |
| Percentage | 6.47% | 42.96% | 44.11% | 6.47% | 100% |
| Mann-Whitney-U Tests (P-Value) | - | 0.891 | 0.104 | 0.539 | - |

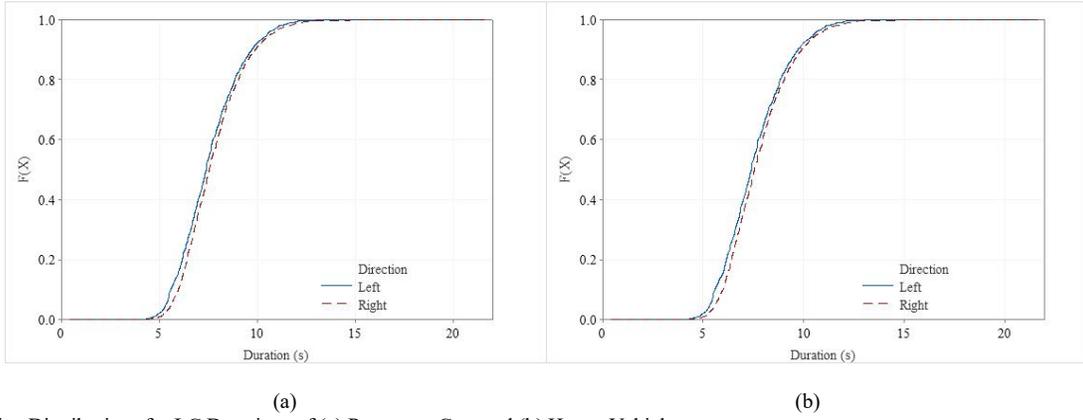

Fig.4. Cumulative Distributions for LC Durations of (a) Passenger Cars and (b) Heavy Vehicles

Durations were extracted using similar methods in previous studies[10, 38]. The start and end time points of a LC are defined as the time when the lateral movement of the subject LC vehicle starts and ends, respectively[32]. The duration is defined as the time from the star to the end points, as shown in Fig.3.

## IV. ANALYSIS OF LC DURATION

### A. Descriptive Analysis of Duration

Table II shows the basic statistics of LC duration for passenger cars and heavy vehicles. Fig.4 shows the cumulative distributions of the duration for passenger cars and heavy vehicles. One can find that the LC durations of heavy vehicles are generally longer than passenger cars, and the difference of the durations between two directions is rather small. To statistically investigate whether there are significant differences in the median values of durations between different LC directions and different vehicle types, Mann-Whitney-U tests were carried out. It is shown in Table III that the average LC duration of heavy vehicles is significantly longer than passenger cars. Results in Table III also show that there is no significant difference between the different directions for heavy vehicles at 0.95 confidence level, while this is not the case for passenger cars. This result is different from previous study based on NGSIM dataset [11], in which the LC direction did not significantly influence the duration. We conjecture that this might be caused by the different traffic condition of the two datasets. Thus, the data were classified into 4 groups according to the subject vehicle speed and the Mann-Whitney-U Tests were carried out separately, as shown in Table IV and Table V.

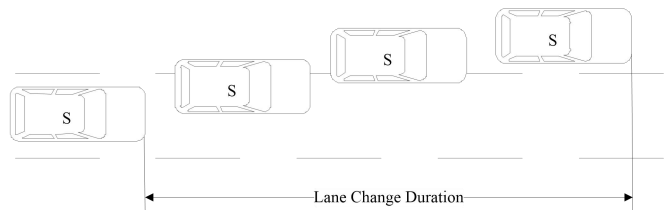

Fig. 3. Definition of LC duration.

One can find that the LC direction does not significantly influence the LC duration within all speed ranges. However, it is interesting to find that the LC direction do have significant influence on LC duration with the speed ranges of [25, 30) m/s and [30, 35) m/s, which is different from previous study based on NGSIM dataset [11]. However, considering the NGSIM data were collected under congestion and transition flow, this finding is not strange. When the traffic is under congestion and transition flow, most passenger cars are pursuing better driving conditions and are restricted by surrounding vehicles. However, when the traffic is under free flow, the motivations of passenger cars are more complicated, which results into the significant difference of the LC durations in different directions. The speed limit in this study is 120 km/h, thus the passenger cars were speeding within the range of [35, 45) m/s , indicating these drivers were more aggressive than other ones and they changed lanes to overtake surrounding vehicles. In this situation, the LC direction did not significantly affect LC durations. To the best of our knowledge, these findings have not been reported in previous studies.

TABLE V
THE NUMBERS OF SAMPLES WITHIN DIFFERENT SPEED RANGES AND RESULTS OF MANN-WHITNEY-U TESTS FOR PASSENGER CARS

| Speed Range (m/s) | [0,20) | [20,25) | [25,30) | [30,35) | [35,45) | Total |
|---|---|---|---|---|---|---|
| Left | 71 | 87 | 440 | 602 | 150 | 1350 |
| Right | 85 | 82 | 254 | 860 | 269 | 1550 |
| Total | 156 | 169 | 694 | 1462 | 419 | 2900 |
| Percentage | 5.38% | 5.83% | 23.93% | 50.41% | 14.45% | 100% |
| Mann-Whitney-U Tests (P-Value) | 0.243 | 0.358 | 0.004 | 0.000 | 0.743 | - |

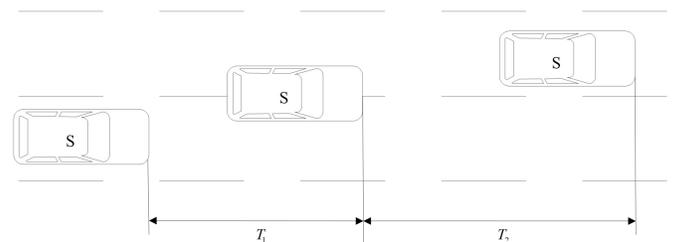

Fig.5. Definition of Different LC Stages

## B. Effect of Speed on LC Duration

It was reported by (Wang et al. 2014) that the navigation speed had significant influence on the LC durations. The LC durations will get saturated when the navigation speed gets higher. To investigate whether this conclusion is still effective under free flow, the effect of navigation speed on LC durations is investigated in this section. Considering the LC direction significantly affects the LC durations of passenger cars, the LCs of passenger cars are classified by directions and speed ranges while those of heavy vehicles are classified only by speed ranges. Table VI shows the statistics of LC durations for passenger cars and heavy vehicles within different speed ranges. Table VII and Table VIII give the results of two-sample Mann-Whitney-U tests to check whether the median value is different in different speed ranges.

From Table VI, one can find that the mean, median values and standard deviations of LC durations for passenger cars generally decrease with the increase of speed. However, the difference of durations within the speed ranges [0, 20) m/s and [20, 25) m/s for LCs to both directions are not large enough. In addition, the median value even increases from 7.280s to 7.340s for LCs to the left when the speed increase from [30, 35) to [35, 45) m/s, indicating the complexity of the effects of navigation speed on LC durations, which can also be confirmed by Table VII. However, it can be observed from Table VII that the LC durations do not significantly vary with navigation speed when it is below 30 m/s. It is interesting to find that the LC duration is easier to be affected by navigation speed when vehicles change lanes to the right. When the navigation speed is higher than 25 m/s, the p-values are all larger than 0.05, indicating the existence of saturated LC duration, which is not observed from LCs to the right.

From Table VI, one can find that the mean and median values of heavy vehicle discretionary LC durations generally decrease with the increase of speed, which is also confirmed by the Mann-Whitney-U tests as most obtained P-values are below 0.05. However, one can also find that the durations for [20, 25) m/s and [25, 30) m/s do not significantly differs with each other as the P-value is 0.778, but the LC duration further decreases for [30, 35) m/s. It means the navigation speed does have significant influence on LC durations of heavy vehicles, and the LC durations of heavy vehicles will reach a saturation value in non-free traffic flow with the increase of navigation speed. However, the LC durations will get a further decease under free flow, which is different from previous study (Wang et al. 2014).

## C. Duration of Different LC Stages

To further investigate the difference of LC durations between different directions, the LC durations are split into two stages. The first stage, $T_1$, is defined as the time from the start point of the LC execution process to the point that the subject vehicle just crosses the line. The second stage, $T_2$, is defined as the time from the end of $T_1$ to the end point of the LC execution process, shown in Fig. 5. Table IX shows the statistics of $T_1$ and $T_2$ for passenger cars and heavy vehicles and Table X shows the results of Mann-Whitney-U tests of $T_1$ and $T_2$ for passenger cars and heavy vehicles.

TABLE VII
RESULTS OF MANN-WHITNEY-U TESTS FOR LC DURATIONS OF PASSENGER CARS WITHIN DIFFERENT VELOCITY RANGES

| LC Direction | Speed Range (m/s) | [20,25) | [25,30) | [30,35) | [35,45) |
|---|---|---|---|---|---|
| Left | [0,20) | 0.754 | 0.174 | 0.033 | 0.029 |
|  | [20,25) | - | 0.041 | 0.003 | 0.004 |
|  | [25,30) | - | - | 0.096 | 0.124 |
|  | [30,35) | - | - | - | 0.677 |
| Left | [0,20) | 0.754 | 0.174 | 0.033 | 0.029 |
|  | [20,25) | - | 0.041 | 0.003 | 0.004 |
|  | [25,30) | - | - | 0.096 | 0.124 |
|  | [30,35) | 3 | 2 | 15 |  |

TABLE VIII
RESULTS OF MANN-WHITNEY-U TESTS FOR LC DURATIONS OF HEAVY VEHICLES WITHIN DIFFERENT VELOCITY RANGES

| Speed Range (m/s) | [0,20) | [20,25) | [25,30) | [30,35) |
|---|---|---|---|---|
| [0,20) | - | 0.049 | 0.020 | 0.000 |
| [20,25) | - | - | 0.778 | 0.008 |
| [25,30) | - | - | - | 0.005 |

TABLE X
RESULTS OF MANN-WHITNEY-U TESTS OF $T_1$ AND $T_2$

| Vehicle Type | Hypothesis Test | Results |
|---|---|---|
| Passenger Cars | Left $T_1$ VS Left $T_2$ | 0.000 |
|  | Right $T_1$ VS Right $T_2$ | 0.348 |
|  | Left $T_1$ VS Right $T_1$ | 0.000 |
|  | Left $T_2$ VS Right $T_2$ | 0.000 |
| Heavy Vehicles | Left $T_1$ VS Left $T_2$ | 0.018 |
|  | Right $T_1$ VS Right $T_2$ | 0.033 |
|  | Left $T_1$ VS Right $T_1$ | 0.310 |
|  | Left $T_2$ VS Right $T_2$ | 0.597 |

TABLE VI
STATISTICS OF LC DURATIONS FOR PASSENGER CARS WITHIN DIFFERENT VELOCITY RANGES

| Vehicle Type | Direction | Speed Range (m/s) | Sample size | Mean(s) | Mean(s) | Median(s) | Std. Dev(s) | Min(s) |
|---|---|---|---|---|---|---|---|---|
| Passenger Cars | Left | [0,20) | 71 | 8.157 | 8.157 | 7.680 | 2.320 | 4.360 |
|  |  | [20,25) | 87 | 8.079 | 8.079 | 7.840 | 1.815 | 4.120 |
|  |  | [25,30) | 440 | 7.633 | 7.633 | 7.480 | 1.6234 | 4.1600 |
|  |  | [30,35) | 602 | 7.442 | 7.442 | 7.280 | 1.4644 | 4.3200 |
|  |  | [35,45) | 150 | 7.352 | 7.352 | 7.340 | 1.351 | 4.120 |
|  | Right | [0,20) | 85 | 8.668 | 8.668 | 8.200 | 2.683 | 4.400 |
|  |  | [20,25) | 82 | 8.566 | 8.566 | 8.180 | 2.328 | 5.080 |
|  |  | [25,30) | 254 | 8.051 | 8.051 | 7.840 | 1.758 | 4.320 |
|  |  | [30,35) | 860 | 7.698 | 7.698 | 7.560 | 1.399 | 4.4800 |
|  |  | [35,45) | 269 | 7.294 | 7.294 | 7.120 | 1.220 | 4.6800 |
| Heavy Vehicles | - | [0,20) | 28 | 9.273 | 9.273 | 9.000 | 1.851 | 6.520 |
|  |  | [20,25) | 127 | 8.565 | 8.565 | 8.120 | 2.053 | 5.120 |
|  |  | [25,30) | 191 | 8.398 | 8.398 | 8.120 | 1.781 | 4.160 |
|  |  | [30,35) | 28 | 7.449 | 7.449 | 7.320 | 1.191 | 5.280 |

For passenger cars, it is interesting to find that $T_1$ and $T_2$ are similar for LCs to the right, but the $T_1$ is much smaller than $T_2$ for LCs to the left, which is also confirmed by the Mann-Whitney-U tests in Table X. It means when change lanes to the left, the drivers of passenger cars tend to take less time to occupy the target lane. Table X also shows that $T_1$ and $T_2$ of LCs to the left are significantly different from those of LCs to the right for passenger cars. This finding indicates that the drivers might use different strategies to perform LCs to different directions, resulting into different LC durations, which means separate LC model according to LC direction is needed for passenger cars.

For heavy vehicles, $T_1$ is a little smaller than $T_2$ for both directions. Table X shows that $T_1$ and $T_2$ do not have significant difference between different directions, indicating the consistence of heavy vehicles' LCs of different directions. Results of Mann-Whitney-U tests show that $T_1$ is significantly shorter than $T_2$ for both directions, which means heavy vehicles generally need more time to adjust their position and speed after encroaching the target lane.

## V. CONCLUDING REMARKS

To investigate the characteristics of durations of discretionary, this study used a dataset containing LC vehicle trajectories of 2905 passenger cars and 433 heavy vehicles.

First, a comprehensive analysis of LC durations was conducted based on vehicle types, LC directions and navigation speeds. It was found that the heavy vehicle takes longer time to complete LC maneuver. The LC direction significantly influences the durations of passenger cars and does not have significant influence on heavy vehicles' LC durations. The navigation speed is found to have an important influence on LC durations. However, it has different impacts according to the vehicle types and LC directions. In general, LC duration of heavy vehicles is more easily affected by navigation speed than passenger cars. For passenger cars, the duration of LC to the right is more easily affected by navigation speed than left LC durations. Further analysis of LC durations at different stages showed that when changing lanes to the left, passenger cars tend to use less time to encroach the target lane while heavy vehicles take more time to adjust their speeds and positions after encroaching the target lanes. However, similar time were used when drivers of passenger cars change lanes to the right. This means that the drivers might use different strategies to perform LCs when they change lanes to different directions. Different from passenger cars, drivers of heavy vehicles used less time to encroach the target lane in both directions.

The analysis confirmed the variation of LC durations according to vehicle types and LC directions. Four models of LC durations were built based on stochastic models. The results of this study can help better understand the mechanism of LCs and the influence of LCs on traffic flow. The proposed LC duration model can improve the accuracy of existing MTSs. However, the data were collected at only one location with 3 lanes. The effects of the road configurations, such as the number of lanes, were not considered in this study. More data will be collected in future to build a general model. In further studies, mandatory LCs will also be investigated to provide more insightful findings.

ACKNOWLEDGMENT

The authors would like to thank Robert Krajewski and Julian Bock (RWTH Aachen Univesity, Aachen, Germany) for providing the highD trajectory data for this study.

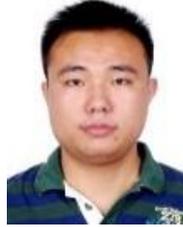

**GEN LI** was born in Yancheng, Jiangsu, China in 1989. He received the bachelor's degree and Ph.D. degree from Southeast University, Nanjing China in 2018. He is currently a lecturer with Nanjing Forestry University. He has published his works in several journals and conferences.

His research interests include microscopic traffic flow theory, traffic simulation, big data and autonomous driving.

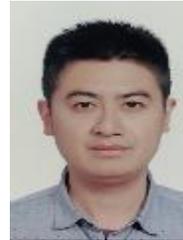

**ZHEN YANG** received the bachelor's degree and Ph.D. degree from Southeast University, Nanjing China in 2009 and 2016. He is currently a lecturer with Nanjing Forestry University. He has published his works in several journals and conferences.

His research interests include microscopic traffic flow theory and ITS.

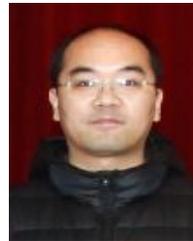

**YIYONG PAN** received his master and Ph.D. degrees from Southeast University, Nanjing, China in 2010 and 2014. He is currently an associate professor with Nanjing Forestry University. He has published his works in several journals and conferences.

His main research areas include the intelligent transportation systems and traffic network theory.